\documentclass[aps, amssymb, amsmath, superscriptaddress, 
twocolumn] {revtex4}
\usepackage{graphicx}
\usepackage{color}
\usepackage{enumitem}
\usepackage{hyperref}

\newcommand{\be}{\begin{equation}}
\newcommand{\ee}{\end{equation}}
\newcommand{\bea}{\begin{eqnarray}}
\newcommand{\eea}{\end{eqnarray}}

\newcommand{\p}{\partial}

\newcommand{\lp}{\left(}
\newcommand{\rp}{\right)}

\renewcommand{\vec}[1]{{\boldsymbol #1}}

\newcommand{\addLL}[1]{\textcolor{red}{#1}}
\definecolor{ForestGreen}{rgb}{0.0,0.68,0.2}
\newcommand{\addMD}[1]{\textcolor{ForestGreen}{#1}}

\begin{document}
\begin{abstract} 
Electron hydrodynamics gives rise to surprising correlated behaviors in which electrons “cooperate” to quench dissipation and reduce the electric fields needed to sustain the flow.  Such collective “free” flows are usually expected at the hydrodynamic lengthscales exceeding the electron-electron scattering mean free path $\ell_{\rm ee}$. Here we predict that in two-dimensional electron gases the collective free flows actually occur at the distances much smaller than $\ell_{\rm ee}$, in a nominally ballistic regime. The sub-$\ell_{\rm ee}$ free flows arise due to retroreflected holes originating from head-on quasiparticle collisions; the holes retrace the paths of impinging electrons and cancel out their potential. An exact solution, obtained in Corbino geometry, predicts potential strongly screened by the hole backflow. Screened potential is described by a fractional power law $r^{-5/3}$ over a wide range of $r$ values, from macroscales down to deep sub-$\ell_{\rm ee}$ scales, and a distinct non-Fermi-liquid temperature dependence.
\end{abstract}

\title{Superscreening by a Retroreflected Hole Backflow 
in Tomographic Electron Fluids}
\author{Qiantan Hong${}^1$, Margarita Davydova${}^1$, Patrick J Ledwith${}^2$, Leonid Levitov}
\affiliation{Physics Department, Massachusetts Institute of Technology, Cambridge MA02139\\
${}^2$Physics Department, Harvard University, Cambridge MA02138}
    
\maketitle
    
 Electron hydrodynamics has emerged recently as a new tool for understanding transport in strongly-interacting electron systems\cite{principi2015,lucas2016b,alekseev2016,guo2017,narozhny2017,scaffidi2017,kashuba2018,lucas2018,derek2018,
 guo2018,kiselev2019a,kiselev2019b,shavit2019}. 
Its appeal stems from the high sensitivity of hydrodynamics to microscopics even in fairly simple Fermi liquids, as well as from anticipation that new kinds of exotic hydrodynamics can arise for exotic quantum matter\cite{crossno2016,bandurin2015,kumar2017,bandurin2018,berdyugin2019,moll2016,baem2018,sulpizio2019,ku2019,jenkins2020}. In this vein, it was predicted recently that two-dimensional electron gases exhibit “tomographic” hydrodynamics, a unique behavior that originates from strong collinear scattering of quasiparticles\cite{ledwith2017,kendrick2018,ledwith2019a,ledwith2019b}. 
These collinear 
scattering processes endow electron fluids with a long-time directional memory that creates an exotic hydrodynamic behavior at large distances, pushing the onset to the conventional Navier-Stokes hydrodynamics to the length scales that greatly exceed the electron 
collision mean free path $\ell_{\rm ee}$. 
    
    \begin{figure}[bt!]
	\includegraphics[width= 0.95\columnwidth]{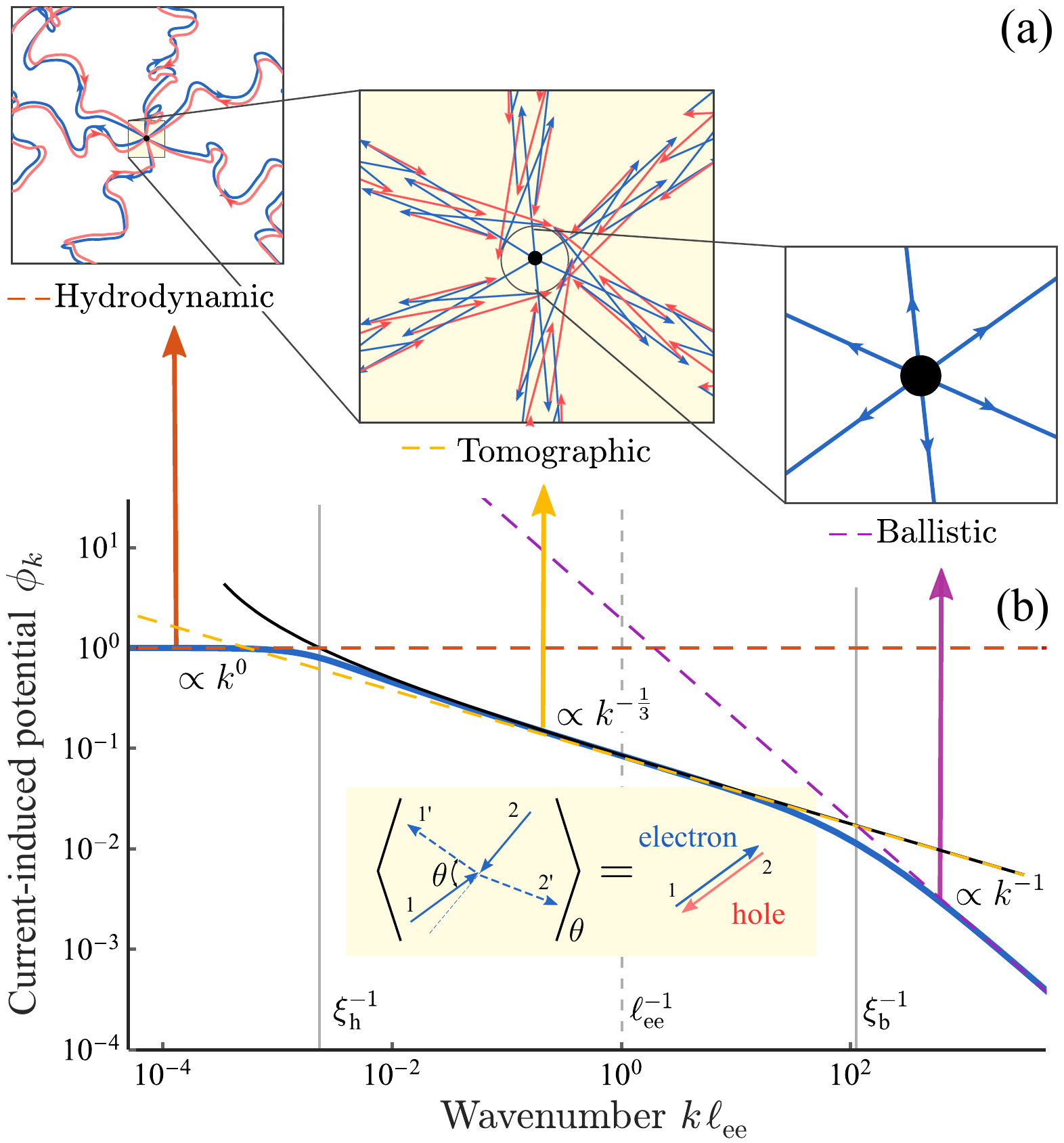}
	\caption{
	(a) Schematic illustration of the spatial hierarchy of the ballistic, tomographic and hydrodynamic regimes for a radial flow. Red and blue lines represent electron trajectories and back-propagating hole trajectories. The tomographic regime (yellow panel) spans a wide range of lengthscales from $r\gg \ell_{\rm ee}$ down to deep sub-$\ell_{\rm ee}$ lengthscales. 
	(b) Current-induced potential distribution around a point source: the exact result obtained from Eqs.\eqref{eq:phi_r},\eqref{eq:Gamma_k} (blue line) and the analytic result for the tomographic regime, Eq.\eqref{eq:s6}, (black line). 
Potential harmonics feature a scaling behavior $\phi_k\sim k^{-\zeta}$ with 
the exponents $\zeta=0$, $1/3$, and $1$ in the three regimes. Dashed lines with these slopes are shown as a guide to the eye. 
	Vertical gray lines mark the boundaries of the tomographic regime $\xi_{\rm h}^{-1}< k < \xi_{\rm b}^{-1}$. The conventional ballistic/hydrodynamic crossover, expected at $k\ell_{\rm ee}=1$ (dashed gray line), does not occur due to the predominance of head-on scattering in 2D electron systems. Parameter values used are given 
	beneath Eq.\eqref{eq:gammas_even_odd}.
	Inset illustrates hole backreflection in head-on scattering, a process that is 
	immune to averaging over scattering angles for the final states. 
	}
	\label{fig1}
	\vspace{-7mm}
	\end{figure}
Here we argue that the tomographic behavior also dominates at ultrashort distances $r\ll \ell_{\rm ee}$, 
overriding  the conventional ballistic regime that normally occurs at such length scales. The unusual behavior at 
sub-$\ell_{\rm ee}$ distances arises due to hole retroreflection. 
The back-reflected holes retrace the paths of impinging electrons, thereby allowing the information about the ee collisions to propagate back into the sub-$\ell_{\rm ee}$ region. As a result, the tomographic dynamics and directional memory effects 
dominate not only at the length scales $r>\ell_{\rm ee}$  but also at $r< \ell_{\rm ee}$, pushing the onset of the ballistic behavior down to abnormally short lengthscales 
    \be
    r \sim \xi_{\rm b}\ll \ell_{\rm ee}, \quad  \xi_{\rm b}/\ell_{\rm ee}\sim \sqrt{T/T_F}\ll 1. 
    \ee
    This peculiar behavior can be understood as a nonclassical quasi-ballistic dynamics of compound objects, the particle-hole pairs composed of particles and holes propagating opposite to each other, as illustrated in Fig.\ref{fig1}. In this regime the electric current is transmitted by a highly coordinated electron flow and a hole counterflow. The 
latter, being equal and opposite to the electron flow, gives a contribution that tends to double the current and cancel out the potential, as illustrated in Fig.\ref{fig1}. 
    Such restructuring of the flow leads to a dramatic reduction (superscreening) of 
    current-induced fields and of Joule dissipation $W\sim \vec j\vec E$, an effect occurring in a wide range of lengthscales, from $r\gg \ell_{\rm ee}$ down to $r\ll \ell_{\rm ee}$.

    From an experimental viewpoint, probing tomographic physics at sub-$\ell_{\rm ee}$ lengthscales has distinct advantages since it considerably softens the limitations that plagued previous searches for electron hydrodynamics. Indeed, the lengthscales at which conventional viscous effects dominate over the ohmic effects satisfy $\ell_{\rm ee} < r < \sqrt{\ell_{\rm ee} \ell_p}$ where $\ell_p$ is the mean free path for momentum-relaxing scattering processes. The materials where the electron fluids are currently being probed, such as graphene or GaAs quantum wells, feature weak phonon and disorder scattering such that the lengthscale $\ell_p$ can exceed $\ell_{\rm ee}$  by as much as an order of magnitude. Yet, the slow square root dependence in $\sqrt{\ell_{\rm ee} \ell_p}$  can make the competing requirements on $r$ challenging to fulfill. 
    
    In contrast, the proliferation of tomographic hydrodynamics to the deep sub-$\ell_{\rm ee}$ lengthscales facilitates probing this physics in experiments. As illustrated in Fig.~\ref{fig1} for the current flowing radially in the Corbino geometry, the counterpropagating electrons and holes comprising the current cancel out each others’ charge; as a result the net potential is suppressed below the values expected for collisionless electron flow. At tomographic lengthscales, the potential drops as a power law $r^{-5/3}$, and then even faster at the length scales where the Navier-Stokes-like viscous regime sets in. Strikingly, this power law behavior, which is shown in Fig.~\ref{fig1}, is identical on both sides of the region $r \sim \ell_{\rm ee}$, where the ballistic/hydrodynamic crossover is usually expected to occur (the line $k\ell_{\rm ee}=1$ in Fig.\ref{fig1}). The suppression of the current-induced potential and the resulting quenching of dissipation upon raising temperature, originating from particle-hole compensation, 
occur at deep sub-$\ell_{\rm ee}$ length scales 
accessible by the state-of-the-art scanning techniques\cite{baem2018,sulpizio2019,ku2019,jenkins2020}. 

A quantity that 
plays the key role in our analysis, yielding a closed-form solution valid in the entire range of relevant spatial scales, is the continued fraction\cite{OS} 
    \be\label{eq:Gamma_k}
    \Gamma(k)=\gamma_2+\frac{z^2}{\gamma_3+\frac{z^2}{\gamma_4+\frac{z^2}{\gamma_5+...}}}
    ,\quad 
    z\equiv kv_F/2
    ,
    \ee
     where $k$ is the wavenumber, related to the spatial scale as $k\sim 1/r$. The quantities $\gamma_m$ are the eigenvalues of the collision operator of 2D electrons, a set of numbers that represent a ``genetic code'' of the 2D electron system giving the relaxation rates for different angular harmonics of the perturbed Fermi surface\cite{ledwith2017,ledwith2019a}. 
     The infinite continued fraction $\Gamma(k)$, defined in the usual way as a limit of finite continued fractions, 
     is well behaved, since the quantities $\gamma_m$ are finite and positive at large $m$.  
     The quantity $\Gamma(k)$ captures all the intricacies of the nonlocal response in the presence of momentum-conserving scattering. 
     
     As a parenthetical remark, other powerful approaches relying on continued fractions have been used recently to tackle various aspects of many-body dynamics\cite{viswanath1994,starykh1997,khait2016,auerbach2019}.
  
    We present a detailed analysis of transport induced by a point current injector, a simple arrangement that mimics Corbino geometry with a rich behavior spanning a wide range of lengthscales as shown in Fig.\ref{fig1}. Fully accounting for the collinear electron scattering, a process that dominates in 2D systems where electron hydrodynamics is currently being probed, our analysis predicts a current-induced potential distribution
    \be\label{eq:phi_r}
     \phi(r)=\int \frac{d^2 k}{(2\pi)^2} e^{i\vec k\vec r} \phi_k
    ,\quad 
     \phi_k=\frac{I}{2\nu e^2 \Gamma(k)}
    ,
    \ee
    where $I$ is the net injected current, $\nu$ is the density of states, $e$ is carrier charge. This result, derived assuming electroneutrality\cite{OS}, is valid at distances greater than the Thomas-Fermi screening length $\lambda_{\rm TF}$. 
    

As a function of $r$ the potential $\phi(r)$ exhibits three different regimes and 
a hierarchy of lengthscales illustrated in Fig.~\ref{fig1}. The tomographic hydrodynamics spans a wide range of scales in between the conventional ballistic and viscous regimes:
\begin{align}\label{eq:xi_b_xi_h}
     \xi_{\rm b} < r <  \xi_{\rm h}
    ,\quad  \xi_{\rm b} \ll \ell_{\rm ee}
    ,\quad  \xi_{\rm h} \gg \ell_{\rm ee}
    ,
\end{align}
pushing the ballistic regime down to deep sub-$\ell_{\rm ee}$ scales $r \sim  \xi_{\rm b}$ and pushing the onset of Navier-Stokes hydrodynamics up to unusually large distances $r \sim  \xi_{\rm h} \gg \ell_{\rm ee}$ [the values $\xi_{\rm b}$ and $\xi_{\rm h}$ are estimated below, see Eqs.\eqref{eq:xi_h_estimate},\eqref{eq:ball-r}].
We predict a power law behavior $\Gamma(k)\sim k^{1/3}$ in the tomographic regime, which translates into a power law dependence of the current-induced potential:\cite{variational_method}
\be\label{eq:phi_r_-53}
\phi(r) \propto I r^{-\frac{5}{3}}
,\quad
  \xi_{\rm b} < r <  \xi_{\rm h}
.
\ee
It is a previously uncharted behavior that is manifested in several surprising effects. 

One is that the small value $\xi_{\rm b}\ll \ell_{\rm ee}$ indicates an expansion of the low-dissipation transport 
to ultrashort distances. 
The origin of this striking behavior, illustrated in Fig.~\ref{fig1}, 
is that the retroreflected holes compensate the charge of the impinging electrons without current relaxation
(since the opposite-moving holes produce the same current as the impinging electrons). As a result, the injected current will flow without 
significantly perturbing the charge and potential distribution in the system. 
    
The large-$r$ behavior is also 
unlike that of classical fluids, where a point injector 
creates pressure gradients and an excess dissipation confined to a thin layer $r\lesssim \ell_{\rm ee}$ near the injector and negligible at larger $r$ \cite{faber_fluid_dynamics,shavit2019}. 
Instead, tomographic hydrodynamics generates a power-law profile extending to much larger distances
$r\sim \xi_{\rm h}\gg \ell_{\rm ee}$.
    
The extinction of electric fields due to hole counterflow resembles some aspects of Andreev hole retroreflection in superconducting NS systems. 
In contrast, the behavior considered here is neither a low-temperature nor a phase-coherent phenomenon, which superconductivity is. In electron fluids it arises at elevated temperatures, becoming prominent at the temperatures for which electron-electron collisions dominate over other collision types. Still, in strong resemblance to Andreev transport, superscreening arises from retroreflected holes which retrace the ballistic paths of impinging electrons.


A useful starting point for our analysis is 
the case when all nonvanishing rates are equal, $\gamma_m=\gamma$, $m\ne 0,\pm1$. 
While it does not describe collinear scattering and tomographic transport, the equal-rate model has been popular in the field since it was introduced in Ref.\cite{molenkamp1995}. In this case the continued fraction is straightforward to evaluate, giving 
$\Gamma(k)=\frac12\lp \gamma+\sqrt{\gamma^2+k^2v^2_F}\rp$. This gives a closed-form expression for the potential distribution
\be\label{eq:gamma'=gamma}
\phi(r)=\frac{I}{2\pi\nu e^2} \int_0^\infty dk \frac{k J_0(kr)}{\gamma+\sqrt{\gamma^2+k^2v^2_F} }
.
\ee
In the absence of scattering, $\gamma=0$, using the identity $\int_0^\infty dx J_0(x)=1$, we recover the $1/r$ profile expected for a radial flow of free particles:
$
\phi(r)=\frac{I}{2\pi \nu e^2 v_F r}
$. 
 In the presence of scattering, $\gamma>0$, the free-particle $1/r$ profile persists up to $r\approx l_{\rm ee}=v_F/\gamma$, dropping sharply to zero 
 at larger $r$. 
 Eq.\eqref{eq:gamma'=gamma} predicts 
 the dependence
 \be\label{eq:phi(r)_exp}
 \phi(r)\approx \frac{I}{2\pi \nu e^2 v_F r} e^{-\lambda r/\xi}
 ,\quad
 \xi=v_F/2\gamma
,
\ee
with a dimensionless $\lambda\approx 1$. 
 The exponential falloff 
at $r>\xi$ marks the onset of the hydrodynamic flow. 
 
In order to 
describe tomographic transport we must account for the effects of collinear collisions. 
In this case the odd-$m$ rates $\gamma_m$ are much smaller than the even-$m$ rates and scale as $m^4$\cite{ledwith2019a,ledwith2019b},  
\be \label{eq:gammas_even_odd}
\gamma_{m\,{\rm even}}=\gamma
,\quad
\gamma_{m\,{\rm odd}}=\gamma' m^4
,\quad m \ll m_{\star}.
\ee
We assume that $\gamma'\ll \gamma$, which describes 
the regime of interest $T\ll \epsilon_F$ (with $\gamma\sim T^2/\epsilon_F$, $\gamma'\sim T^4/\epsilon_F^3$). 
The odd-$m$ rates $\gamma_m$ initially grow as $m^4$, saturating at the even-$m$ value $\gamma$ at a large $m \gtrsim m_{\star}= (\gamma/\gamma')^{1/4}$. 

The current-induced potential,  
Eq.\eqref{eq:phi_r}, is illustrated in Fig.~\ref{fig1}(b) for the ratio $\gamma'/\gamma=5 \times 10^{-8}$, with 
the wavenumber measured in units of $\ell_{\rm ee}^{-1}=\gamma/v_F$ 
[the details of evaluating continued fractions are given in \cite{OS}]. 
A small value $\gamma'/\gamma$ was chosen to 
widen the tomographic regime to illustrate small deviations from scaling discussed below. 

The predicted dependence $\phi_k$ asymptotes to a constant at small $k$ and to $1/k$ at large $k$. This checks with $\phi(r)$ falling off abruptly at large distances and behaving as $1/r$ at short distances, similar to the conventional 
transport, 
Eq.\eqref{eq:phi(r)_exp}. 
A new, tomographic regime with a power-law scaling $\phi_k\sim k^{-1/3}$ occurs at intermediate $k$ values, reflecting the behavior at the lengthscales where transport is governed by collinear collisions. Importantly, the new regime extends to abnormally large distances $r\gg \ell_{\rm ee}$ and starts at ultrashort sub-ballistic distances $r\ll \ell_{\rm ee}$. This is a signature of particle-hole compensation due to hole retroreflection that tends to screen the current-induced potential all the way back to the source.

To demystify the origin of the extremely short screening length 
we note that
the backreflection of the hole is misaligned from the direction of the outgoing electron by a small angle $\delta \theta \sim m_*^{-1} \sim 
(T/\epsilon_F)^{1/2}\sim (\gamma'/\gamma)^{1/4}
$
\cite{ledwith2017}. After a collision, the hole will return on average to a point in space offset by the distance $\xi_{\rm b}\approx \ell_{\rm ee}\delta \theta \sim 
\ell_{\rm ee}(\gamma'/\gamma)^{1/4}
$ from the electron source. This is illustrated in the middle panel of Fig.\ref{fig1}(a): the holes flow 
outside the circle $r\approx\xi_{\rm b}$.  This estimate coincides with the ballistic-tomographic crossover length  $\xi_{\rm b}$ found below, Eq.~\eqref{eq:ball-r}.


    

To gain insight into the scaling regimes pictured in 
Fig.\ref{fig1}, we develop an analytic approach which yields a closed-form result for $\phi_k$ and establishes the exact value of the scaling exponent. This will be done through analyzing the behavior of continued fractions $\Gamma(k)$ vs. $k$. 

First, having in mind that $k$ values and distances are related as $r\sim 1/k$, we expand Eq.\eqref{eq:Gamma_k} in small $k$. This gives $\Gamma(k)=\gamma_2+
k^2v_F^2/4 \gamma_3
+O(k^4)$. Eq.\eqref{eq:phi_r} then predicts potential decaying at large $r$ as $\phi(r) \sim  r^{-1/2}e^{-r/\xi_{\rm h}}$ with 
\be\label{eq:xi_h_estimate}
\xi_{\rm h} = \frac{v_F}{2 \sqrt{\gamma_2 \gamma_3}}\propto \frac{\epsilon_F}{T}\ell_{\rm ee}
\ee 
a lengthscale that can be identified with the onset of Navier-Stokes hydrodynamics. As discussed above, the abnormally large value $\xi_{\rm h}$ reflects proliferation of the tomographic regime to large distances.

At large $k$, in contrast, one has to analyze 
the whole continued fraction, taking the limit $\gamma_m\ll |z|$. 
Despite this being a subtle limit to take, the end result is 
easy to understand from the selfconsistent relation 
\be
\label{eq:Gamma_delta}
\Gamma(k)=\lim_{\gamma_m\ll z}\lp \gamma_2+\frac{|z|^2}{\gamma_3+\frac{|z|^2}{\gamma_4+\cdots}}\rp =\frac{|z|^2}{\Gamma(k)}
\ee
This relation predicts $\Gamma(k)=|z|=\frac{kv_F}{2}$. 
Eq.\eqref{eq:phi_r} then 
yields the potential 
that matches our expectation for a ballistic flow near the source:
\begin{equation}\label{eq:ball-r}
\phi(r)=\frac{I}{2\pi\nu e^2 v_F r}
,\quad
r< \xi_{\rm b} =3 \frac{v_F}{\gamma} \left( \frac{\gamma'}{\gamma}\right) ^{\frac{1}{4}} \ll \ell_{\rm ee}
.
\end{equation}
A surprising finding here is the ultrashort range of distances where the ballistic flow occurs. While a superficial inspection of Eq.\eqref{eq:Gamma_delta} might suggest that the scale $\xi_{\rm b}$ coincides with $\ell_{\rm ee}=v_F/\gamma$, a correct estimate which accounts for a large number of terms in the continued fraction predicts abnormally short values 
$\xi_{\rm b}\ll \ell_{\rm ee}$. As discussed above, this indicates that the ballistic behavior is largely overridden by the tomographic effects.

To derive the scaling behavior in the tomographic regime, we 
start with the following observation. 
In general, $\gamma_m$ in Eq.\eqref{eq:s0.5} depends on $m$, with large differences between successive even and odd $m$. 
To capture this behavior in a simplified model, we set $\gamma_m=\gamma_{\rm e}$ for all even $m$ and $\gamma_{\rm o}$ for all odd $m$, ignoring the $m$ dependence of $\gamma_{\rm e}$ and $\gamma_{\rm o}$. 
In this case, $\Gamma(k)$ can be evaluated exactly: 
\begin{equation}
\Gamma(k)=
\frac12\sqrt{\frac{\gamma_{\rm e}}{\gamma_{\rm o}}}
\lp \sqrt{\gamma_{\rm o}\gamma_{\rm e}}+\sqrt{\gamma_{\rm o}\gamma_{\rm e}+k^2v_F^2}\rp
\end{equation}
This motivates introducing ``level-$m$'' partial continued fractions, defined as
\begin{equation}\label{eq:s0.5}
\Gamma_m(k)=\gamma_m+\frac{|z|^2}{\gamma_{m+1}+\frac{|z|^2}{\gamma_{m+2}+\cdots}}
.
\end{equation}
These quantities can be evaluated similarly, giving
\begin{equation}
\Gamma_m(k)=\frac{b_m}{2}\lp \sqrt{\gamma_{\rm o}\gamma_{\rm e}}+\sqrt{\gamma_{\rm o}\gamma_{\rm e}+k^2v_F^2}\rp
\end{equation}
where $b_m = \sqrt{\gamma_{\rm e}/\gamma_{\rm o}}$ for even $m$ and $b_m = \sqrt{\gamma_{\rm o}/\gamma_{\rm e}}$ for odd $m$. 
When the even/odd parity separation of $\gamma_m$ is significant, $\gamma_{\rm e}\gg \gamma_{\rm o}$, the quantities $\Gamma_m$ are much larger for even $m$ than for odd $m$.
However, despite this even/odd beating effect, $\Gamma_m$ remains nearly constant for each individual parity of $m$. 

Based on these observations, we expect that for a realistic low-temperature model with $\gamma_{\rm e}(m)$ and $\gamma_{\rm o}(m)$ slowly varying with $m$, $\Gamma_m$ for each individual parity will also be slowly varying with $m$. It is then natural to analyze the dependence $\Gamma_m$ vs. $m$ for a fixed parity. 
The quantity $\Gamma(k)$ will then be found by taking $m=2$.

We therefore proceed to construct a recursion relation that connects $\Gamma_m$ and $\Gamma_{m+2}$. Taking a difference and using Eq.\ref{eq:s0.5} yields
\begin{equation}
  \label{eq:s1}
  \Gamma_{m}-\Gamma_{m+2}=\gamma_m-\frac{\gamma_{m+1}\Gamma_{m+2}^2}{\gamma_{m+1}\Gamma_{m+2}+|z|^2}
\end{equation}
It turns out, perhaps surprisingly, that this nonlinear relation is greatly simplified after the substitution
\begin{eqnarray}\label{eq:s1.1}
\Gamma_m = \frac{|z|^2}{ \gamma_{m-1}}\left(\frac{u_{m}}{u_{m+2}}-1\right)
,
\end{eqnarray} 
which transforms it into a linear relation
\begin{align}\nonumber 
\frac{|z|^2}{\gamma_{m+1}}u_{m+4}+\frac{|z|^2}{\gamma_{m-1}}u_m
=\left(\gamma_m+\frac{|z|^2}{\gamma_{m-1}}+\frac{|z|^2}{\gamma_{m+1}}\right)u_{m+2}
. 
\end{align}
By regrouping the terms, 
the relation above can be  
cast into the form resembling 
a discretized second-order ODE: 
\begin{align}\label{eq:s1.3}
&  \frac{1}{2}\left(\frac{|z|^2}{\gamma_{m-1}}+\frac{|z|^2}{\gamma_{m+1}}\right)(u_{m+4}-2u_{m+2}+u_{m})
\\ \notag
& -\frac{1}{2}\left(\frac{|z|^2}{\gamma_{m-1}}-\frac{|z|^2}{\gamma_{m+1}}\right)(u_{m+4}-u_{m})=\gamma_{m}u_{m+2}
.
\end{align}
We emphasize that these relations are totally general. 
Indeed, our starting point is a tridiagonal system of equations for the amplitudes of different harmonics\cite{OS}. In this case there is a natural bipartite structure (the off-diagonal couplings are between harmonics of different parity). Eliminating variables of one parity yields a tridiagonal problem describing the other parity. 
Hereafter we take $m$ values 
to be even. 


Since $\Gamma_m$ and $\gamma_m$, when restricted to a fixed parity, are both slowly varying with $m$, we take Eq.\eqref{eq:s1.3} to continuous domain by replacing the differences with derivatives 
\begin{eqnarray}
4 \frac{|z|^2}{\gamma_{\rm o}} \frac{d^2u}{dm^2}-4 |z|^2 \frac{d \gamma_{\rm o}^{-1}}{dm} \frac{du}{dm}=\gamma_{\rm e} u\notag
.
\end{eqnarray}
where $\gamma_{\rm e}(m) = \gamma_m$ and $\gamma_{\rm o}(m) = \gamma_{m+1}$. This simplifies to
\begin{eqnarray}\label{eq:s2}
u''-\frac{\gamma_{\rm o}'}{\gamma_{\rm o}}u'-\frac{\gamma_{\rm o}\gamma_{\rm e}}{4|z|^2}u=0
.
\end{eqnarray}
In the continuous domain, Eq.\eqref{eq:s1.1} now reads
\begin{eqnarray}\label{eq:s3}
\Gamma(m) = -\frac{2|z|^2}{ \gamma_{\rm o} u}\frac{du}{dm}
.
\end{eqnarray}
We assume that $\gamma_{\rm o}$ and $\gamma_{\rm e}$ both converge to a constant value $\gamma$ for high harmonics, $m\gtrsim m_{\star}$, then the values $\Gamma_m$ for $m\gtrsim m_{\star}$ are given by the solution of the one-rate model $\frac{1}{2}\lp\gamma+\sqrt{\gamma^2+k^2v_F^2}\rp$. Therefore, we can write the boundary condition for Eq.\eqref{eq:s2} as
\begin{equation}\label{eq:s1.5}
 \left.-\frac{2|z|^2}{ \gamma u}\frac{du}{dm}\right|_{m=m_{\star}}=\Gamma(m_{\star})=\frac{\gamma+\sqrt{\gamma^2+k^2v_F^2}}{2}
 ,
\end{equation}
and proceed to solve Eq.\eqref{eq:s2} on the interval $0<m<m_{\star}$ to obtain the value of $\Gamma_{m=2}$.

From now on we focus on the $m^4$ model, Eq.\eqref{eq:gammas_even_odd}, 
wherein $\gamma_{\rm o} = \gamma' m^4$, $\gamma_{\rm e} = \gamma$. 
Then Eq.\eqref{eq:s2} becomes
\begin{eqnarray}\label{eq:s3.5}
u''-\frac{4}{m}u'-\frac{\sqrt{\gamma\gamma'}}{4|z|^2}m^4u=0
\end{eqnarray}
This equation, after introducing a new variable
$ g=\frac{\sqrt{\gamma\gamma'}}{6|z|}m^3
$
and replacing $u$ with $w=u/g^{5/6}$, is transformed into the Bessel equation for $w(g)$, yielding a general solution 
\begin{equation}\label{eq:s4} 
u=g^{5/6}[C_1 I_{5/6}(g)
+ C_2I_{-5/6}(g)]
\end{equation}
where $I_{\alpha}(g)$ is the $\alpha$th-order modified Bessel function of the first kind. 

From Eq.\eqref{eq:s3}, we know that constant prefactor of $u$ does not affect $\Gamma(m)$ values. 
Thus the only quantity that remains to be determined is the ratio $C_2/C_1$. 
When $m$ varies from $0$ to $m_{\star}= (\gamma/\gamma')^{1/4}$, $g$ varies from $0$ to $g_{\star} = \gamma^{5/4}/(6\gamma'^{1/4}|z|)$, and the boundary condition for $u(g)$ at the right end of the interval is given by Eq.\eqref{eq:s1.5}
\begin{equation}\label{eq:s4.5}
  -\frac{m_{\star}\gamma}{6g_{\star}u}\left.\frac{du}{dg}\right|_{g=g_{\star}} =\Gamma(m_{\star})>0
\end{equation}
When $k \ll \frac{\gamma^{5/4}}{3v_F \gamma'^{1/4}}\equiv \xi_{\rm b}^{-1}$, Eq.\eqref{eq:s4.5} requires $\Gamma > 0$ for $g_{\star} \gg 1$.
However, the large-argument asymptotic expansions for $I_\alpha(g)$ are exponentially growing in exactly the same way for $\alpha$ and $-\alpha$.
This results in a monotonically increasing $u$, 
producing unphysical values $\Gamma(m_\star)<0$.
The only way to resolve this conflict is to pick $C_2\simeq -C_1$,
then all terms in the asymptotic expansion cancel out.


We therefore arrive at Eq.\eqref{eq:s4} with $C_1=-C_2$. 
The quantity 
of interest, $\Gamma(k)$, corresponds to 
$g|_{m=2} = \frac{4}{3} \frac{\sqrt{\gamma\gamma'}}{|z|}$.
Since for $k \gg  \xi_{\rm h}^{-1} \sim \frac{\sqrt{\gamma\gamma'}}{v_F}$, the latter quantity is small, $g|_{m=2} \ll 1$, and we can expand $u$ in $g\ll 1$\cite{Gamma_vs_Gamma}:
\begin{equation}\label{eq:u_asymptotic}
  u=-\frac{2^{\frac{5}{6}}}{\Gamma(\frac{1}{6})}+\frac{2^{\frac{5}{6}}}{\Gamma(\frac{11}{6})}\lp\frac{g}{2}\rp^{5/3}-\frac{1}{2^{\frac{1}{6}}\Gamma(\frac{7}{6})}\lp\frac{g}{2}\rp^{2}+\cdots
.
\end{equation}
Plugging it in Eq.\eqref{eq:s3} 
gives the dependence 
\begin{equation}\label{eq:s6}
  \Gamma(k) = 
  \frac{\Gamma(\frac{1}{6})}{3^{\frac{2}{3}}2^{\frac{4}{3}}\Gamma(\frac{5}{6})}\frac{\gamma}{(\gamma\gamma')^{\frac{1}{6}}}
  |z|^{\frac13}-2^{-\frac23}\gamma
, 
\end{equation}
valid provided $\xi_{\rm h}^{-1}\ll k\ll\xi_{\rm b}^{-1}$, which is exactly the condition for the tomographic regime.
For such $k$ the first term in Eq.\eqref{eq:s6} dominates, giving 
the power-law dependence $\Gamma_{m=2} \propto k^{1/3}$. 
This dependence, combined with Eq.\eqref{eq:phi_r}, yields a $r^{-5/3}$ scaling for the current-induced potential profile, Eq.\eqref {eq:phi_r_-53}. 

When the separation between the lengthscales $\xi_{\rm b}$ and $\xi_{\rm h}$ becomes smaller (i.e., $\gamma'/\gamma$ becomes larger), the second (subleading) term in Eq.\eqref{eq:s6} cannot be
neglected anymore, producing a change in the apparent scaling exponent values, $\Gamma(k)\propto k^{\alpha}$ and $\phi(r)\propto r^{-2+\alpha}$, $\alpha>1/3$.



Conveniently, the crossover to the ballistic dependence occurs at abnormally short sub-$\ell_{\rm ee}$ distances, where the $r^{-5/3}$ potential is strong, lending 
tomographic behavior amenable to state-of-the-art scanning probe techniques\cite{baem2018,sulpizio2019,ku2019,jenkins2020}. 
%
Besides the short lengthscales, 
detection of tomographic transport is facilitated by the unique temperature dependence. Indeed, the e-e collision rate responsible for the backflow of holes will grow with temperature and, accordingly, 
potential in the system will diminish; the temeprature-induced suppression will occur over a wide range of length scales, including sub-$\ell_{\rm ee}$ length scales. Simultaneously the lengthscale $ \xi_{\rm b}$ where the tomographic behavior sets in will become shorter as temperature grows, providing clear a experimental signature of superscreening. 



We thank E. I. Rashba for insightful discussions of the Thomas-Fermi screening effects in an out-of-equilibrium  current-carrying state of the Fermi gas\cite{OS}. 
This work was supported by the Science and Technology Center for Integrated Quantum Materials, NSF Grant No. DMR-1231319; Army Research Office Grant W911NF-18-1-0116; and Bose Foundation Research fellowship.

\begin{thebibliography}{99}
\bibitem{OS} The continued fraction structure is a direct consequence of transport equations assuming a form of a tridiagonal matrix in the angular harmonics basis, see Appendix.
\bibitem{Gamma_vs_Gamma} The Gamma functions in Eqs.\eqref{eq:u_asymptotic} and \eqref{eq:s6} are not to be confused with the notation for the continued fractions $\Gamma(k)$.
\end{thebibliography}

\begin{appendix}
\section{Continued fractions formalism for a point source}

\subsection{Transport equation}
    

To model the behavior of space charge in the presence of quasiparticle scattering, we will assume a weak deviation from equilibrium and use the linearized transport equation with the collision term $I_{\rm ee}$ describing two-body scattering: 
\be\label{eq:r,t}
(\p_t+\vec v\cdot\vec \nabla_r 
-I_{\rm ee})\delta f(\vec p,\vec r) +e\vec E\cdot\vec v\frac{\p f_{\rm eq}}{\p\epsilon}=s_{\vec p,\vec r}
.
\ee
The source term  $s_{\vec p,\vec r}=s_0\delta(\vec r)(-\p f_{\rm eq}/\p\epsilon)$ represents an injector, here without loss of generality placed at the origin. The electric field is 
due to the space charge induced by the flow, 
$\vec E(\vec r)=-\vec \nabla \delta\phi(\vec r)$, with the space-charge potential 
\be\label{eq:phi_n}
\delta\phi(\vec r)=\int d^2 r'\frac{e\delta n(\vec r')}{\kappa |\vec r-\vec r'|}
,\quad
\delta n(\vec r)=
\sum_{\vec p}\delta f(\vec p,\vec r)
,
\ee 
where $\kappa$ is the effective dielectric constant. 
The solution of this 
problem will describe restructuring of the flow, ballistic near the source and tomographic at larger distances (Fig.\ref{fig1}). It will also account for the non-local field-charge response, Eq.\ref{eq:phi_n}, through the change in $\vec E$ due to a current-induced deviation in the carrier distribution from equilibrium, as well as the Thomas-Fermi screening. 

It will be convenient to transform our problem to an auxiliary easier-to-solve problem for a fictitious free-particle distribution $\tilde f(\vec p,\vec r)$ obtained by selfconsistently shifting the local chemical potential by an amount that depends on the local carrier depletion. 
The new problem, described by Eq.\eqref{eq:r,t} with $\vec E=0$, will then be solved in a closed form in the Fourier representation. 
%
%

We first rewrite Eq.\eqref{eq:r,t} by taking the perturbed distribution and its potential to be a plane wave 
\be\label{eq:plane_wave}
\delta f(\vec p,\vec r)=\delta f_{\vec k}(\vec p) e^{i\vec k\vec r-i\omega t}
, \quad
\delta \phi(\vec r)=\delta \phi_{\vec k} e^{i\vec k\vec r-i\omega t}
,
\ee
to obtain 
\be\label{eq:wk}
(-i\omega+i\vec k\cdot\vec v-I_{\rm ee})\delta f_{\vec k}(\vec p)
-i\vec k\cdot\vec v e\delta \phi_{\vec k}  \frac{\p f_{\rm eq}}{\p\epsilon}=s_{\vec p,\vec r}
\ee
where $e\delta\phi_{\vec k}=U(k)\delta n_{\vec k}$ are harmonics of the current-induced space-charge potential, with  $U(k)=\frac{2\pi e^2}{\kappa k}$ the 2D Coulomb potential formfactor. 

In the regime of interest, $T\ll E_F$, the perturbed distribution $\delta f$ is concentrated near the Fermi level and can be 
represented by angular harmonics describing the Fermi surface modulation evolving in space and time, 
\be\label{eq:harmonics}
\delta f_{\vec k}(\vec p)=-\frac{\p f_{\rm eq}(p)}{\p\epsilon}\sum_{m=-\infty}^\infty 
\delta f_m e^{im\theta}
\ee
where $\theta$ is the angle parameter on the Fermi surface. For conciseness, we will suppress the dependence of the harmonics $\delta f_m$ on the wavenumber $\vec k$, Eq.\eqref{eq:plane_wave}, restoring it at the end. 
The factorization into the radial and angular dependence described by $-\frac{\p f_{\rm eq}(p)}{\p\epsilon}$ and 
the sum of harmonics $\delta f_m e^{im\theta}$, respectively, is an approximation that captures the behavior of the low-lying excitations in a Fermi gas at $T\ll E_F$. 

Because of the cylindrical symmetry, the collision operator is diagonal in the $e^{im\theta}$ basis, 
\be
I_{\rm ee} e^{im\theta} =-\gamma_m e^{im\theta}
,
\ee
with the eigenvalues $\gamma_m$ 
describing the relaxation rates for different angular harmonics of the perturbed distribution. 
Different values $\gamma_m$ 
account for different scattering processes in the system. 
Here we analyze the 
two-rate model\cite{ledwith2019a,ledwith2019b} in which the odd-$m$ rates $\gamma_m$ are much smaller than the even-$m$ rates at small enough $m$, and scale as $m^4$. As $m$ grows the odd-$m$ and even-$m$ rates eventually become equal. The dependence $\gamma_m$ vs. $m$ can be described as 
\be \label{eq:gammas_even_odd_with_cutoff}
\gamma_{m\,{\rm even}}=\gamma
,\quad
\gamma_{m\,{\rm odd}}=\frac1{\frac1{\gamma' m^4}+\frac1{\gamma}}
.
\ee
The parameter values of interest at temperatures $T\ll\epsilon_F$ correspond to $\gamma'\sim T^4/\epsilon_F^3$ much smaller than $\gamma\sim T^2/\epsilon_F$ and, in addition, $\gamma_{m=0}=\gamma_{m=\pm1}=0$ for the zero-mode harmonics. 
The crossover value  $m$ above which the even-$m$ and odd-$m$ rates become approximately equal,
\be
m=m_{\star}=(\gamma/\gamma')^{1/4}
,
\ee
grows as $\sqrt{\epsilon_F/T}$ with temperature decreasing, $T\ll\epsilon_F$. 

\subsection{Computing the response function}

A transformation to an auxiliary problem for a fictitious free-particle distribution 
can now be achieved as follows. 
We first note that the field term in Eq.\eqref{eq:wk} is a product of the $p$ harmonic $i\vec k\vec v$ and an angle-independent function $\delta \phi_{\vec k} \frac{\p f_{\rm eq}}{\p\epsilon}$ that depends on the injected current. This structure can be exploited to absorb the field term into the streaming term $i\vec k\vec v \delta f_{\vec k}(\vec p)$ by introducing an auxiliary distribution function $\delta\tilde f$
for which the $m=0$ harmonic is rescaled by the  
dielectric function 
\be\label{eq:epsilon_k}
\epsilon_k=1+\nu U(k)
\ee
with $\nu$ the density of states at the Fermi level, whereas other harmonics remain unchanged:
\be
\delta \tilde f_{m=0}=\epsilon_k \delta f_{m=0}
,\quad
\delta \tilde f_{m\ne 0}=\delta f_{m\ne 0}
.
\ee
The new distribution $\delta \tilde f$ obeys Eq.\eqref{eq:wk} with $\delta\phi_{\vec k}=0$; the potential of the space charge is given in terms of $\delta \tilde f$ by the relation
\be\label{eq:phi_space_charge_1}
\phi(\vec r)=\sum_{\vec k} e^{i\vec k\vec r} \frac{U(k)}{\epsilon_k} 
\nu \delta \tilde f_{m=0}(\vec k)
,
\ee
where the density of states arises in the usual manner by approximating the sum over the states near the Fermi level as $\sum_{\vec p} -\frac{\p f_{\rm eq}(p)}{\p\epsilon}=\nu$. 
The fictitious particle distribution $\delta {\tilde f}$ obeys the transport equation, Eq.\eqref{eq:wk}, in which the space charge potential is suppressed: 
\be\label{eq:wk_tilde_f}
(-i\omega+i\vec k\cdot\vec v-I_{\rm ee})\delta \tilde f_{\vec k}(\vec p)
=-s_0 \frac{\p f_{\rm eq}(p)}{\p\epsilon} 
\ee
A general solution of Eq.\eqref{eq:wk_tilde_f} can be given in terms of continued fractions. For that we exploit the structure of the streaming term $i\vec v\vec k\delta f$ in Eq.\eqref{eq:wk_tilde_f} which, in the angular harmonics basis, represents a nearest-neighbor ``hopping'' that couples harmonics $m$ and $m\pm1$. This observation allows us to rewrite Eq.\eqref{eq:wk_tilde_f} 
as system of coupled algebraic equations for the Fourier coefficients $\delta \tilde f_m$ as follows:
\be\label{eq:tight-binding}
(\gamma_m-i\omega)\delta \tilde f_m+
iz\delta \tilde f_{m+1}+
i\bar z \delta \tilde f_{m-1}=s_0\delta_{m,0}
\ee
Here we defined a complex parameter $z=\frac{v}2(k_x+ ik_y)$, 
and used the identity 
$\vec k\vec v=
z e^{-i\theta}+\bar z e^{i\theta}$.

The coupled equations in Eq.\eqref{eq:tight-binding} can be solved recursively as follows. 
For $m> 0$ we define the quantities $\alpha_m=i\delta \tilde f_{m+1}/\delta \tilde f_{m}$; 
in terms of $\alpha_m$ the $m>0$ equations read 
\be
\gamma_m+z\alpha_{m}-\frac{\bar z}{\alpha_{m-1}}=0
,
\ee 
where from now on, for brevity, we suppress $i\omega$. 
These relations can be transformed to a recursion relation $\alpha_{m-1}=\frac{\bar z}{\gamma_m+z\alpha_{m}}$ and iterated over $m+1$, $m+2$, ..., to obtain
\be
\alpha_{m-1} 
=\frac{\bar z}{\gamma_m+\frac{|z|^2}{\gamma_{m+1}+\frac{|z|^2}{\gamma_{m+2}+...}}}
\ee
Similarly, for $m<0$ we define the quantities $\beta_m=i\delta \tilde f_{m-1}/\delta \tilde f_{m}$; in terms of $\beta_m$ the $m<0$  equations read
\be
\gamma_m-\frac{z}{\beta_{m+1}}+\bar z\beta_{m}=0
,
\ee 
In this case, expressing $\beta_{m+1}$ through $\beta_m$ as $\beta_{m+1}=\frac{z}{\gamma_m+\bar z\beta_m}$ and iterating over $m-1$, $m-2$, ...,  yields
\be
\beta_{m+1} = 
\frac{z}{\gamma_m+\frac{|z|^2}{\gamma_{m-1}+\frac{|z|^2}{\gamma_{m-2}+...}}}
.
\ee
We can now find the harmonic $\delta \tilde f_0$ from the $m=0$ equation in which we set $\gamma_0=0$,
\be
iz\delta \tilde f_1+i\bar z \delta \tilde  f_{-1}=s_0
.
\ee 
Writing $\delta \tilde f_1=-i\delta \tilde f_0\alpha_0$ and $\delta \tilde f_{-1}=-i\delta \tilde f_0\beta_0$, 
substituting the continued fraction representation for $\alpha_0$ and $\beta_0$ and setting $\gamma_1=\gamma_{-1}=0$, yields a relation for the zeroth harmonic which describes the space charge density: 
\be
\delta \tilde f_0 =\frac{s_0}{2\Gamma(k)}
,\quad
\Gamma(k)=\gamma_2+\frac{|z|^2}{\gamma_3+\frac{|z|^2}{\gamma_4+...}}
\ee 
The continued fraction $\Gamma(k)$ is well behaved at $\omega=0$, since the quantities $\gamma_m$ are finite and positive at large $m$. The $\omega$ dependence, which can be obtained by an analytic continuation from small $\omega$ values, will be discussed elsewhere.  

Using the relation between the physical and fictitious $m=0$ harmonics, $\delta f_0=\frac1{\epsilon_k}\delta \tilde f_0 = \frac1{1+\nu U(k)}\delta \tilde f_0$, $U(k)=\frac{2\pi e^2}{\kappa k}$, we can write a closed-form expression for the harmonics of the potential
\begin{align}
\phi(k)=\frac{\nu U(k)}{1+\nu U(k)}\delta \tilde f_0
\end{align}
Combining this relation with the above result for $\delta \tilde f_0$ and linking the source term value to the injected current, 
\be
s_0=\frac{I}{e\nu}
,
\ee
yields the representation of the current-induced potential in terms of the continued fraction $\Gamma(k)$, Eqs.\eqref{eq:phi_r},\eqref{eq:Gamma_k}, which is exploited in the main text. 

In agreement with the charge neutrality requirement, 
this nonlocal relation turns into a local relation at distances larger than the Thomas-Fermi screening length, $r>\lambda_{\rm TF}=\kappa/2\pi e^2\nu$, giving
\be\label{eq:phi_space_charge_2}
\phi(\vec r)=
\delta \tilde f_{m=0}(\vec r)
.
\ee
This transformation, which replaces the actual distribution $f$ with a fictitious distribution $\tilde f$ obeing the free-particle problem, provides a general recipe to analyze the space charge buildup induced by currents in a nonequilibrium system. 
Indeed, the net space charge density can be expressed through the $m=0$ harmonic as 
\be
\delta n_{\vec k}=\nu e\delta f_{m=0}
=\frac{\nu e}{\epsilon_k}\delta \tilde f_{m=0}
.
\label{nk}
\ee
This relation between the actual current-induced density change and the fictitious free-particle density buildup can be viewed as an extension of the Thomas-Fermi mean-field screening theory to a non-equilibrium transport problem; as such it is valid at first order in current.  

The behavior of the potential 
is illustrated in Fig.~\ref{fig1}(b). The ratio of $\gamma'/\gamma$ that was used is $5 \times 10^{-8}$ and the wavenumber is measured in units of $\ell_{ee}$. The infinite continued fraction is computed by setting a large threshold value of $m$ after which $\gamma_{m\, {\rm odd}} = \gamma_{m\, {\rm even}} = \gamma$ and the rest of the continued fraction is given by the explicit expression for the one-rate model.  A very small value of $\gamma'/\gamma$ was chosen to enlarge the range of lengthscales spanned by the tomographic regime -- more than four decades in Fig.~\ref{fig1}(b) -- and exhibit the small deviations from scaling discussed in the main text. 

\end{appendix}
\end{document}